\documentclass[letterpaper,twocolumn,10pt]{article}
\usepackage{usenix-2020-09}

\usepackage{tikz}
\usepackage{amsmath}
\usepackage{comment}
\usepackage{subcaption}

\begin{document}

\date{}

\title{\Large \bf Prospects for Improving Password Selection}

\author{
{\rm Eryn Ma }\\
Pomona College
\and
{\rm Summer Hasama}\\
Pomona College
\and
{\rm Eshaan Lumba}\\
Pomona College
\and
{\rm Eleanor Birrell}\\
Pomona College
\and
{\rm Emails: \{ymac2020,shaa2020,elaa2020\}@mymail.pomona.edu, eleanor.birrell@pomona.edu }
} 

\maketitle

\begin{abstract}
User-chosen passwords remain essential to online security, and yet people continue to choose weak, insecure passwords. In this work, we investigate whether \emph{prospect theory}, a behavioral  model of how people evaluate risk, can provide insights into how users choose passwords and whether it can motivate new designs for password selection mechanisms that will nudge users to select stronger passwords. We ran a user study with 762 participants, and we found that an intervention guided by prospect theory---which leverages the reference-dependence effect by framing selecting weak passwords as a loss relative to choosing a stronger password---causes approximately 25\% of users to improve the strength of their password (significantly more than alternative interventions) and reduced the final number of weak passwords by approximately 25\%. We also evaluate the relation between user behavior and users' mental models of hacking and password attacks. These results provide guidance for designing and implementing account registration mechanisms that will significantly improve the strength of user-selected passwords, thereby leveraging insights from prospect theory to improve the security of systems that use password-based authentication. 
\end{abstract}

\section{Introduction}

User-chosen passwords remain a critical component of security. Many efforts have been made to nudge users towards choosing stronger passwords, including password rules~\cite{kelley2012guess} and password meters~\cite{egelman2013does}, but these efforts have met with only partial success. Password recipes are ineffective at enforcing strong password choices~\cite{weir2010testing,kelley2012guess}, many password meters are ineffective~\cite{carnavalet2015large} especially for accounts users consider unimportant~\cite{egelman2013does}, and  users continue to select and use weak passwords~\cite{passwords2020}. In this work, we investigate the extent to which insights from cognitive psychology apply to users' password selection decisions and how those insights might be leveraged to enhance security by nudging users to select stronger passwords. 

\emph{Prospect theory}~\cite{kahneman1979prospect,tversky1981framing,tversky1986framing,tversky1991loss,tversky1992advances} is an experimentally-grounded behavioral model of how people make decisions under risk. 
Prospect theory has been successfully applied to various different areas of economics, and it has proven a useful model both for  explaining observed behaviors~\cite{shefrin1985disposition,meng2018can,kHoszegi2007reference,barseghyan2013nature,kHoszegi2009reference,heidhues2014regular,camerer1997labor,crawford2011new,snowberg2010explaining} and for prescriptively nudging people towards higher-utility choices~\cite{fryer2012enhancing,levitt2016behavioralist,thaler2004save,hossain2012behavioralist}. 

Interactions between humans and systems that affect security and privacy can be framed as decisions under risk. For example, password selection requires users to evaluate the risk associated with each possible password they consider (how likely is it that their account will be compromised if they select that password, and how bad will the consequences be if that occurs) and balance that risk against other competing factors (e.g., memorability and easy of typing, including on mobile devices) in order to decide which password to use. However, prior work has thus far explored the intersection between prospect theory and security and privacy only in limited specific domains, such as investment in security~\cite{verendel2008prospect,schroeder2005using}, adoption of two-factor authentication~\cite{qu2019towards}, and disclosure of personal information~\cite{grossklags200725,adjerid2013sleights,acquisti2013privacy}.

In this work, we explore the connection between two effects identified in the prospect theory literature---the \emph{reference-dependence effect} and the  \emph{source-dependence effect}---and password selection. We ran a user study with 762 participants in which we asked people to create an account on an example website. Users who initially selected weak passwords or moderate passwords were presented with an interactive prompt asking whether they wanted to go back and choose a stronger password; there were six different versions of the interactive prompt corresponding to three different framings (positive, neutral, and negative) and two different prompt phrasings (specific and vague). Participants also completed a follow-up survey about their beliefs regarding  passwords and hacking threats.


To investigate the effect of individual design factors, we evaluated whether users who initially selected weak password ultimately improve the strength of their chosen password.  We found that the reference-dependence effect applies to password selection decisions---i.e., an interaction with negative framing resulted in significantly higher rates of improvement compared to neutral framing ($p<.001$) or positive framing ($p=.022$). However, the source-dependence effect did not appear to apply; the phrasing of the prompt (specific of vague) did not have a significant impact on whether user went back and selected a stronger password ($p=.611$). 

To validate the impact of a prospect-driven intervention, we evaluated whether our intervention with negative framing increased the strength of user-selected passwords. We found that the final number of weak passwords (after interacting with the negative framing) was significantly lower than than the initial number of weak passwords selected by our participants ($p=.019$). This results confirm that the reference-dependence effect can be leveraged to enhance security by nudging users to select stronger passwords. 

Finally, we investigated whether mental models of security affected how users responded to our interactions. We found that perceptions about who is likely to be targeted by hackers are correlated with password selection decisions. However, password selection decisions were consistent across different mental models of risks associated with password attacks. 

Our results suggest that prospect theory can be a helpful model for understanding how users make security decisions---in particular, how users choose passwords. We find that an interaction after users select a preliminary password that leverages a negative framing---i.e., frames continuing with weak passwords as having a negative impact on security relative to improving the password before creating an account---can significantly strengthen passwords and reduce the number of weak passwords that users ultimately select. We believe that this insight from prospect theory can form the foundation for designing and implementing  password selection mechanisms that enhance security by nudging users to make better password selection  decisions.

\section{Background: Prospect Theory}

Prospect theory~\cite{kahneman1979prospect,tversky1981framing,tversky1986framing,tversky1991loss,tversky1992advances}---first introduced in the 1970s as a critique of the then-dominant expected utility theory~\cite{vonNeumann1944theory,friedman1948utility}---is a descriptive model of decision making in the presence of risk. Expected utility theory---which asserts that a principal faced with a choice between two options will evaluate the expected utility of each outcome and then select the option with the higher expected utility---does not accurately predict human behavior observed in many experimental settings.

\begin{figure}[t!]
	\begin{center}
		\begin{subfigure}[b]{.48\columnwidth}
			\includegraphics[width = \textwidth]{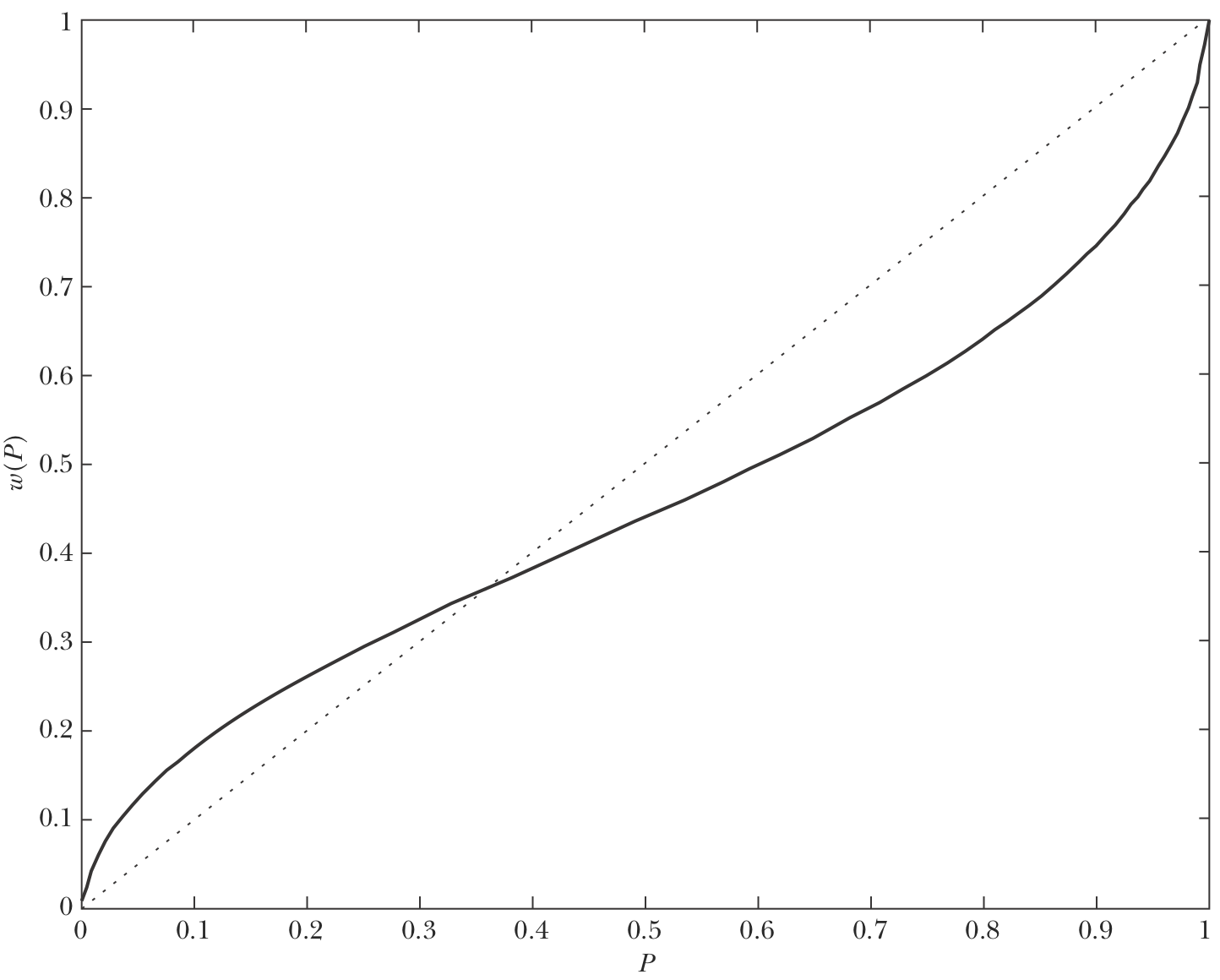}
			\caption{Decision weight function $w$}
			\label{fig:weight_function}
		\end{subfigure}
		\begin{subfigure}[b]{.48\columnwidth}
			\includegraphics[width = \textwidth]{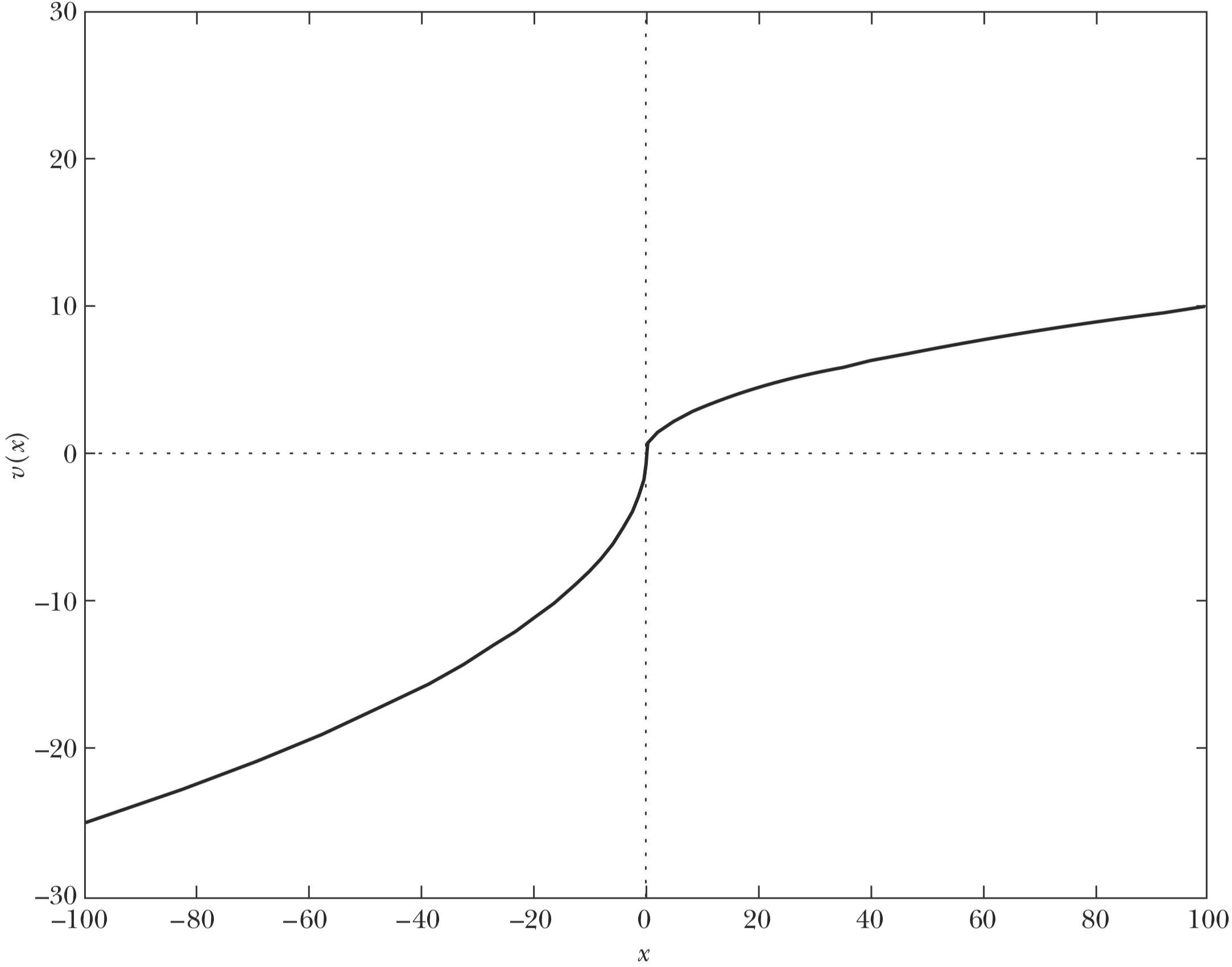}
			\caption{Subjective value function $\nu$}
			\label{fig:value_function}
		\end{subfigure}
	\end{center}
	\caption{Example functions matching empirically-observed behavior proposed by \cite{tversky1992advances}.} 
	\label{fig:example_functions}
\end{figure}

Prospect theory instead posits that decisions are comprised of two phases: an editing phase and an evaluation phase. In the editing phase, humans apply a set of simplifying heuristics to reduce the complexity of the decision problem. In the evaluation phase, probabilities and utilities are weighted by a decision weight $w$ and a subjective value $\nu$, respectively; example functions capturing empirically-observed behavior are shown in Figure~\ref{fig:example_functions}. 
Humans are then presumed to rationally evaluate the options based on the weighted expected subjective value of the edited prospects. 

The interactions between the editing phase and the weighting functions $w$ and $\nu$ result in several effects that have been empirically validated through a series of experimental studies:

\begin{enumerate}
	\item \textit{Isolation Effect:} People simplify decision problems by disregarding components shared between alternatives and focusing exclusively on components that distinguish the options. 
	\item \textit{Pseudocertainty Effect:} People simplify decision problems by treating extremely likely (but uncertain) outcomes as though they were certain. 
	\item \textit{Reference-dependence Effect:} People simplify decision problems by defining outcomes relative to a neutral baseline. The framing of a problem can effect which baseline is used. 
	\item \textit{Certainty Effect:} People overweight the probability of outcomes that are certain relative to outcomes that are merely probable.
	\item \textit{Source-dependence Effect:} People have different decision weights depending on the type of risk. For example, people have higher decision weights for contingent risks than for equivalent probabilistic risks (e.g., they prefer an insurance policy that provides certain coverage of specific types of damages to one that provides probabilistic coverage of all types of damages). Similarly, people are \emph{ambiguity averse}---they prefer to bet based on precisely defined odds rather than on unknown odds. 
	\item \textit{Loss Aversion Effect:} People subjectively dislike losses more than they value gains. That is, the value function is steeper for negative values (losses) than for positive values (gains). 
\end{enumerate}

\begin{figure*}[t!]
	\begin{center}
		\begin{subfigure}[b]{.48\textwidth}
			\includegraphics[width = \textwidth]{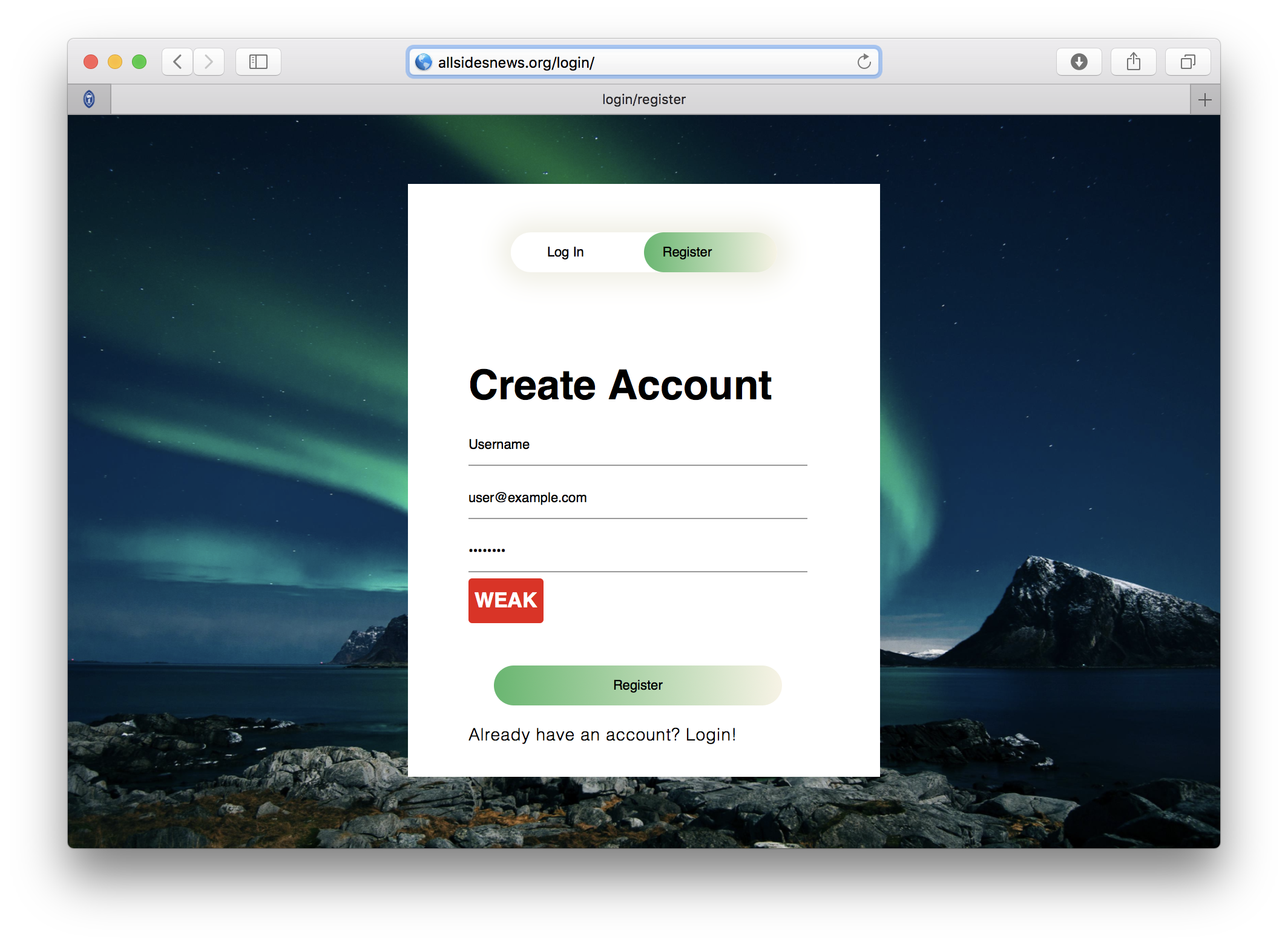}
			\caption{Account creation page}
			\label{fig:login}
		\end{subfigure}
		\begin{subfigure}[b]{.48\textwidth}
			\includegraphics[width = \textwidth]{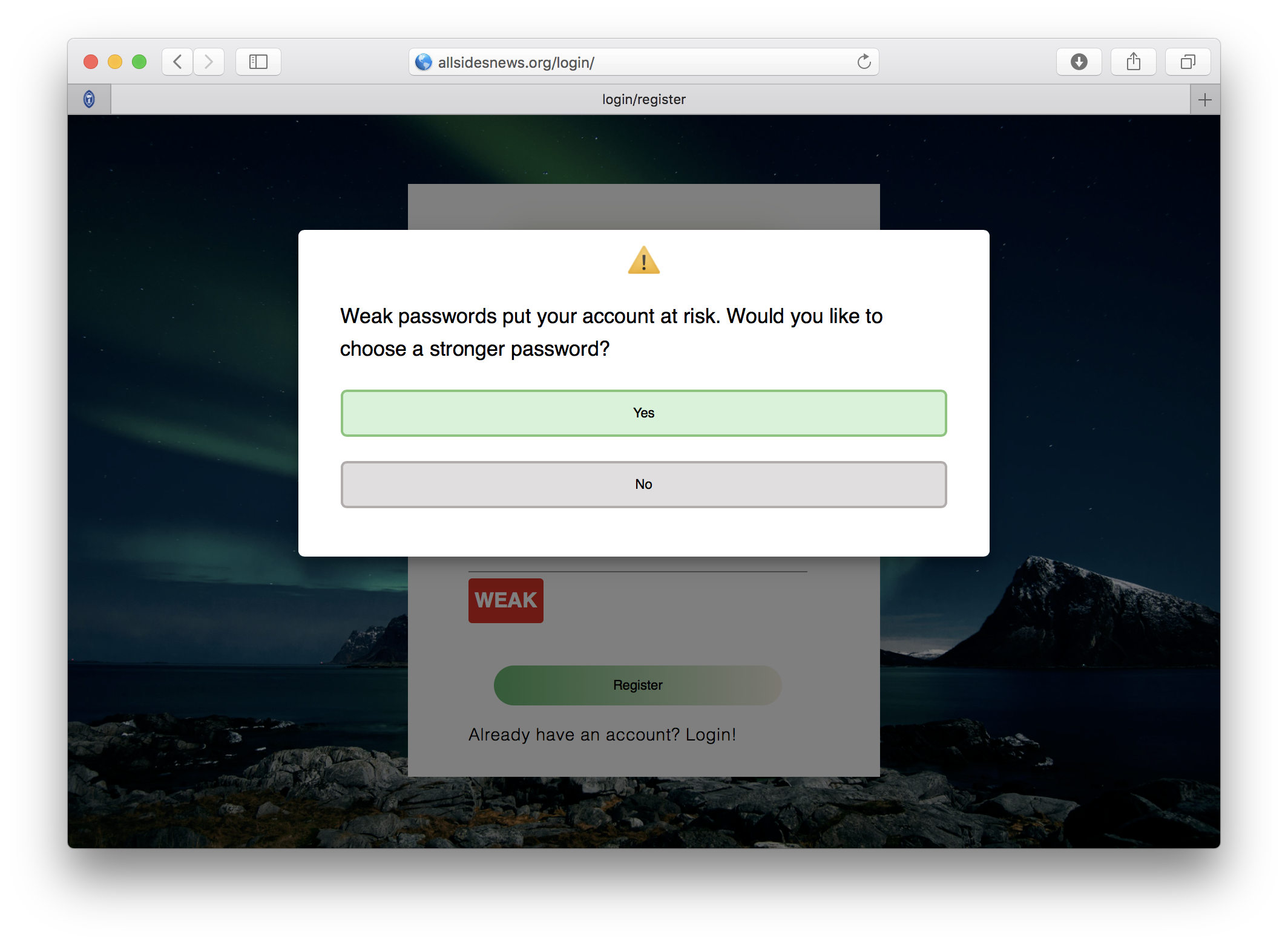}
			\caption{Example interactive prompt}
			\label{fig:example-popup}
		\end{subfigure}\\
		\begin{subfigure}[b]{.48\textwidth}
			\includegraphics[width = \textwidth]{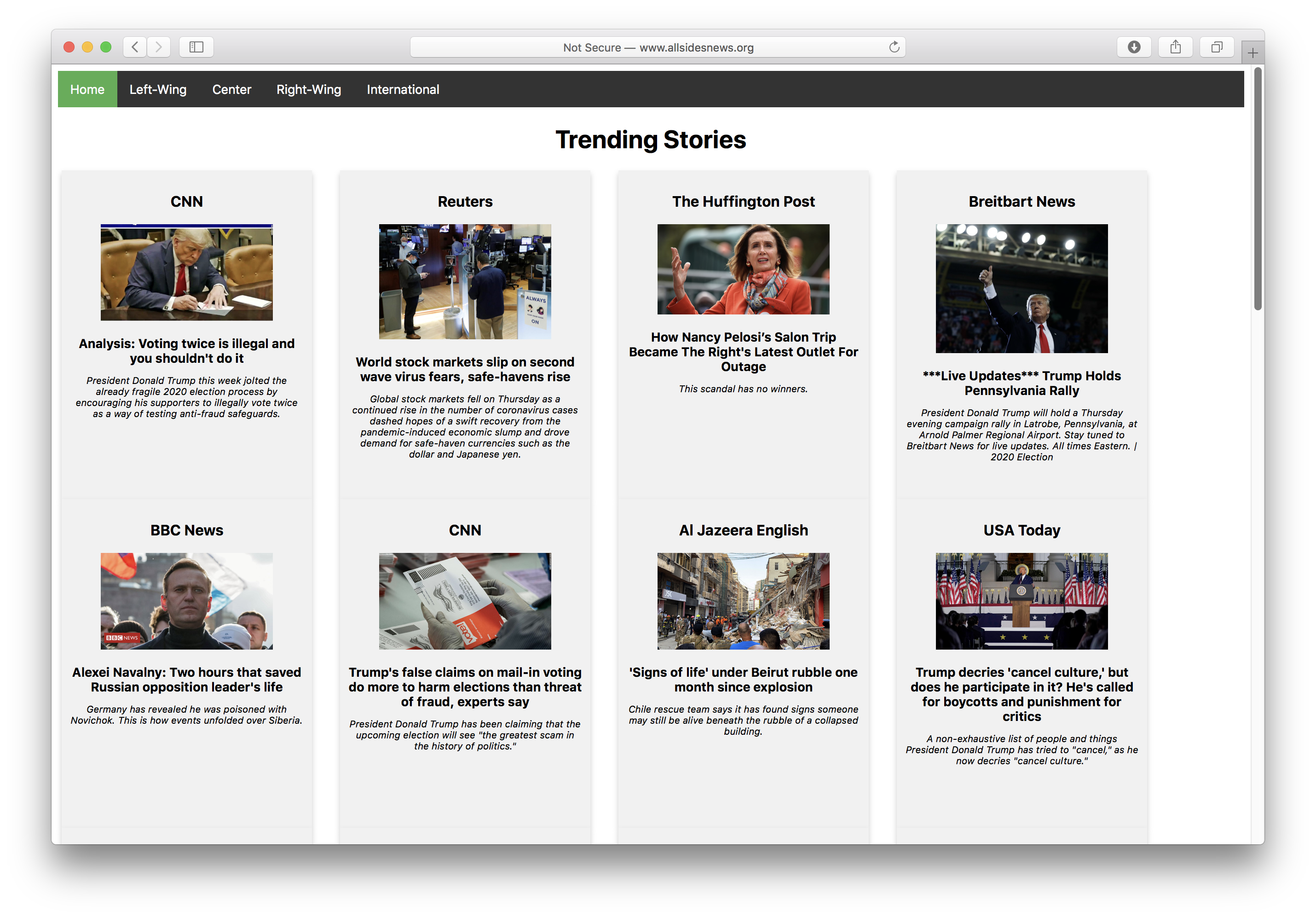}
			\caption{Website home page}
			\label{fig:homepage}
		\end{subfigure}
	\end{center}
	\caption{Screenshots of the account creation process on the example site}\label{fig:screenshots}
\end{figure*}

More than 40 years later, prospect theory is still widely viewed as the best available model for how people make decisions in the presence of risk. It has been applied as a descriptive model to explain observed behavior in various different areas of economics including finance~\cite{shefrin1985disposition,barberis2008stocks,dimmock2010loss,meng2018can}, insurance~\cite{kHoszegi2007reference,sydnor2010over,barseghyan2013nature,hu2007behavioral}, savings~\cite{kHoszegi2009reference}, price setting~\cite{heidhues2014regular}, labor supply~\cite{camerer1997labor,crawford2011new}, and betting markets~\cite{snowberg2010explaining}. Within the domain of computer science, prospect theory has been applied to explain decisions relating to investment in security~\cite{verendel2008prospect,schroeder2005using}, adoption of two-factor authentication~\cite{qu2019towards}, and disclosure of personal information~\cite{grossklags200725,acquisti2013privacy,adjerid2013sleights}.

Prospect theory has also been applied prescriptively in certain domains to nudge people towards certain ``desirable'' behaviors, including nudging employees to increase their retirement contributions~\cite{thaler2004save}, encouraging teachers to improve student outcomes~\cite{fryer2012enhancing}, and incentivizing teams in high-tech factories to increase their productivity~\cite{levitt2016behavioralist}. 
However, prospect-driven interventions have not been uniformly successful: a 2012 study did not see any increase in effort when financial or non-financial incentives for students were framed as losses compared to equivalent incentives framed as gains~\cite{hossain2012behavioralist}.

\section{Methodology}

To investigate how well prospect theory effects apply as a descriptive model of password selection, we conducted a user study ($n=762$) to evaluate the impact of two prospect theory effects---the source-dependence effect and the reference-dependence effect---on password selection decisions.

\subsection{Experimental Setup}

We developed an experimental aggregated news site that is accessible only to authenticated users. When visiting for the first time, a user is redirected to the account creation page (Figure~\ref{fig:login}). During account creation, the password strength is classified in real time using the \texttt{zxcvbn} password strength estimator~\cite{wheeler2016zxcvbn} as weak, moderate, or strong, and this information is displayed to the user in real-time via a three-level password meter. A password is classified as weak if the it has  a \texttt{zxcvbn} total score of 0 or 1 and moderate if the password has a  total score of 2. Passwords with a total score of 3 or 4 are considered strong.

\begin{figure*}
	\centering
	\begin{subfigure}[b]{.32\textwidth}
		\includegraphics[width=\textwidth]{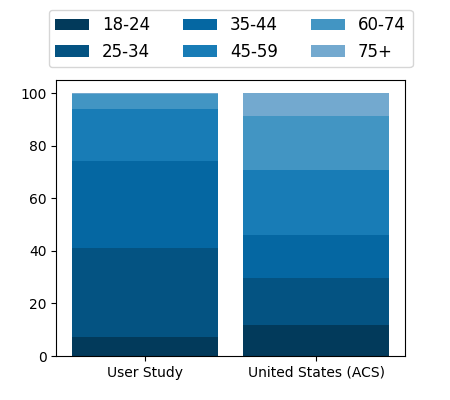}
		\caption{Age}
		\label{fig:demographics-age}
	\end{subfigure}
	\centering
	\begin{subfigure}[b]{.32\textwidth}
		\includegraphics[width=\textwidth]{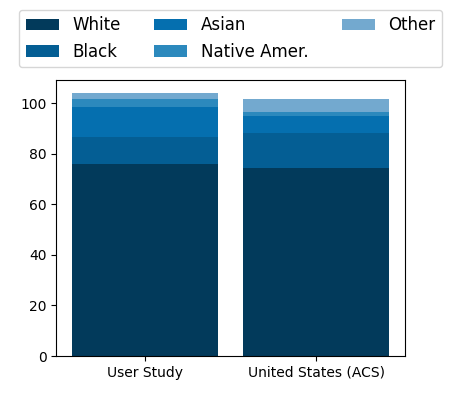}
		\caption{Race}
		\label{fig:demographics-race}
	\end{subfigure}
	\centering
	\begin{subfigure}[b]{.3\textwidth}
		\includegraphics[width=\textwidth]{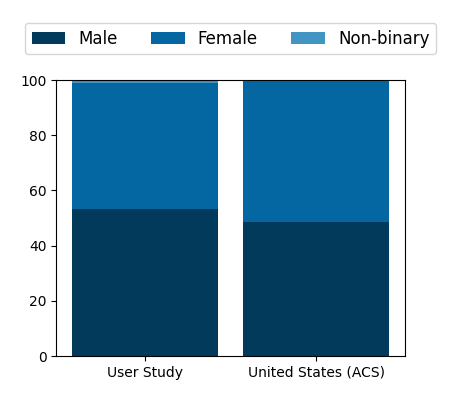}
		\caption{Gender}
		\label{fig:demographics-gender}
	\end{subfigure}
	\caption{Comparison between the demographics of our study participants and the demographics of the United States, as published in the American Community Survey (ACS).}
	\label{fig:demographics}
\end{figure*}

After initially selecting a password, users who select a strong password are redirected to the home page of the aggregated news site (Figure~\ref{fig:homepage}). Users who select a weak or moderate password are instead presented with an interactive prompt that states that weak (resp., moderate) passwords put their account at risk and asks whether they would like to choose a stronger password (Figure~\ref{fig:example-popup}). This prompt was presented using one of two possible wordings:
\begin{enumerate}
	\item \textit{Vague Prompt:} The password you selected is $\langle\textit{strength}\rangle$. Would you like to choose a stronger password?
	\item \textit{Specific Prompt:} $\langle\textit{strength}\rangle$ passwords can be guessed or learned by attackers in $\langle\textit{time}\rangle$, which may lead to the loss of personal information, including credit card info, and identity theft. Would you like to choose a stronger password?
\end{enumerate}
Here $\langle\textit{strength}\rangle$ is the classification based on the \texttt{zxcvbn} total score of the password the user submitted: weak or moderate. (Recall that users who submit a strong password are authenticated immediately and are not presented with a prompt.) 
For the specific prompt, $\langle\textit{time}\rangle$ is the \texttt{zxcvbn} estimate for how long it would take to crack the password with an offline guessing attack if passwords are hashed and salted using a slow hashing algorithm with a moderate work factor (e.g., bcrypt, scrypt,  or  PBKDF2). 

As shown in Figure~\ref{fig:example-popup}, the prompt has two buttons: one to go back (and choose a different password) and one to continue creating the account with the current password. This pair of buttons is labeled with one of three possible framings: 
\begin{enumerate}
	\item \textit{Positive Framing:}
	\\ \textbf{Go Back:} Choose a stronger password to reduce the risks of financial loss and identity theft
	\\ \textbf{Continue:} Create account with current password
	\item \textit{Neutral Framing:}
	\\ \textbf{Go Back:} Yes
	\\ \textbf{Continue:} No
	\item \textit{Negative Framing:}
	\\ \textbf{Go Back:} Choose a stronger password
	\\ \textbf{Continue:} Ignore potential risks of financial loss and identity theft and create account with current password
\end{enumerate}
Users who elect to continue are redirected to the site homepage. Users who choose to go back stay on the account creation page until they select and submit a second password; they are then redirected to the site home page (no matter how strong their second selected password is). 

Each user is pseudorandomly assigned to one of the six conditions---corresponding to the cross product of the two prompts and the three framings---based on a hash of their current IP address. For each user, the site logs the strength of their initial password choice, how they interact with the interactive prompt (for users who initially select a weak or moderate password), and the strength of their revised password choice (for users who go back to change their password).

\subsection{User Study}

To evaluate the impact of prospect theory effects on password selection decisions, we conducted a user study with 762 users. Participants were recruited through Amazon Mechanical Turk, and participation was restricted to United States residents who had completed at least 50 HITs with an approval rate of at least 95\%.  This user study, including all consent forms and survey instruments, was reviewed in advance by the Institutional Review Board (IRB) at our institution and received an IRB exemption approval.

The task was advertised as beta-testing an aggregated news site. Each participant was asked (1) spend 1-2 minutes exploring the website as they would normally behave as an Internet user, (2) enter the unique confirmation code displayed when they visited the site, and (3) complete the follow-up survey questions. Participants who did not enter a valid confirmation code were excluded from the study.  

Prior to the start of the task, users were presented with a consent form that informed them about what data would be collected and how that data would be used; only people who consented to these practices participated in the study. To avoid the appearance of collecting any personal information, users were given a username and email address to use on the site. 

The follow-up survey contained questions about participants' mental model of hacking~\cite{wash2010folk} and their views on password security. We also asked basic demographic questions.  To ensure validity, we conducted a series of three cognitive interviews and made revisions to clarify the wording of our survey questions. A copy of our final survey instrument is included in Appendix~\ref{appendix:survey}. 

Participants who completed the task and entered a valid confirmation code were compensated \$1.20. The median completion time for the full task was 5.15 minutes.

A summary of the demographics of our user study population, along with a comparison to the overall demographics of the United States is shown in Figure~\ref{fig:demographics}. 
We observe briefly that the demographics of our study population differ slightly from the overall demographics of the united states. In particular, our survey population skews significantly younger and slightly more male. Black and African American demographics are slightly underrepresented while Asain and Asian American demographics are slighly overrepresented. Despite these differences, studies conducted with Mechanical Turkers about security and privacy have generally been found to extend to the broader population~\cite{redmiles2019well}.

\subsection{Hypotheses}

To explore the impact of prospect-driven interventions on password selection decisions, we identified and evaluated four hypotheses relating to the source-dependence effect, the reference-dependence effect, and users' mental models.

\paragraph{Source-dependence effect.} When presented with the vague interactive follow-up prompt, a user is required to evaluate options in the presence of multiple different sources of risk: in addition to reasoning about how likely it is that an attacker would target this site and/or this user, the user must evaluate uncertainties about how hard it would be for an attacker to guess their password and about what the potential consequence of password compromise might be. When presented with the specific prompt, some of these uncertainties---in particular how hard it would be for an attacker to guess their password and what an attacker might do after they've learned a user's password---are eliminated in favor of more concrete risks. 

The source-dependence effect observes that users evaluate different types of risk differently, and in particular that ambiguities are evaluated differently than more concrete risks. We therefore hypothesize that users will evaluate the the option to continue with their current (weak or moderate) password more negatively when presented with the specific prompt than with the vague prompt, resulting in stronger password selection after interacting with the specific prompt compared to the vague prompt. \\

\textit{\textbf{Hypothesis 1:} Users' password selection decisions exhibit the \emph{source-dependence effect}, that is users assigned to the specific prompt conditions are more likely to strengthen their password and less likely to ultimately select a weak password compared to users assigned to the vague prompt conditions.} 

\paragraph{Reference-dependence effect.} In the conditions with positive framing, the option to go back is labeled as ``Choose a stronger password to reduce the risks of financial loss and identity theft''. By emphasizing the benefits of going back, this framing implicitly nudges the user to consider the option to continue as the neutral reference point and the option to go back as a choice with higher utility relative to that reference point. By contrast, the negative framing emphasizes the loss of utility (``potential risks of financial loss and identity theft'') associated with continuing with the current password, thereby implicitly nudging the user to treat the option to go back as the neutral reference point and to evaluate continuing as a loss of utility relative to that reference point.

The reference-dependence effect implies that this difference in framing will cause users assigned to a positive framing condition to evaluate the difference between going back (i.e., choosing a stronger password) and continuing (i.e., submitting a weak password) as a positive \emph{gain} in utility, whereas users assigned to a negative framing condition will evaluate the difference between continuing (i.e., submitting a weak password) and going back (i.e., choosing a stronger password) as a \emph{loss} of utility. The loss aversion effect suggests that the subjective value function is steeper for (relative) losses than for (relative) gains. We therefore hypothesize that users will evaluate the option to continue with the current (weak) password more negatively in the negative framing conditions than the positive framing conditions---even though the two options have the same absolute utility in all conditions---resulting in better password selection decisions after interacting with the negative framing prompt compared to the neutral and positive framing prompts. \\

\textit{\textbf{Hypothesis 2:}  Users' password selection decisions  exhibit the \emph{reference-dependence effect}, that is users assigned to a negative framing condition---which frames going back as the neutral baseline and continuing as a loss relative to that baseline---are more likely to strengthen their password and less likely to ultimately select a weak password compared to users assigned to  vague prompt conditions.} 

\paragraph{Mental Models of Hacking.} Our user study concluded with a series of questions about participants' mental models of hacking and password security. One question we asked was who participants believe were the primary target of password stealing attacks. Drawing on Wash's taxonomy of hacker mental models~\cite{wash2010folk}, we provided three possible answer: hackers target everyone equally, hackers primarily target rich people, and hackers primarily target users with special privileges (e.g., system administrators). We hypothesized that users who believe that hackers target everyone equally will consider themselves to be a more likely target compared to users with other mental models and will therefore be more sensitive to risks associated with password compromise.\\

\textit{\textbf{Hypothesis 3:} Users who believe everyone is equally likely to be targeted by a password stealing attack will be more likely to strengthen their password after seeing an interactive prompt and will be more likely to initially choose a strong password. } \\

We also asked participants questions designed to understand how their evaluation of password risks. First we asked how likely they believed a password stealing attack would be to succeed if they selected a weak (resp., moderate, strong) password. We also asked users to select which things a hacker would be able to do if they successfully compromised a users password, given a list of six possibilities. We hypothesized that users' beliefs about risks associated with passwords would correlate with users' password selection decisions.\\

\textit{\textbf{Hypothesis 4:} Users who believe that weak passwords are more likely to be guessed by attackers or who believe that there are more potential consequences when a password gets compromised will be more likely to strengthen their password after seeing an interactive prompt and will be more likely to initially choose a strong password. }

\section{Results}

To evaluate our hypotheses, we observed which passwords users ultimately selected (after interacting with the prompt) and compared those final password selection decisions to users' initial password selections.


\begin{figure}[t!]
	\begin{center}
		\includegraphics[width = \columnwidth]{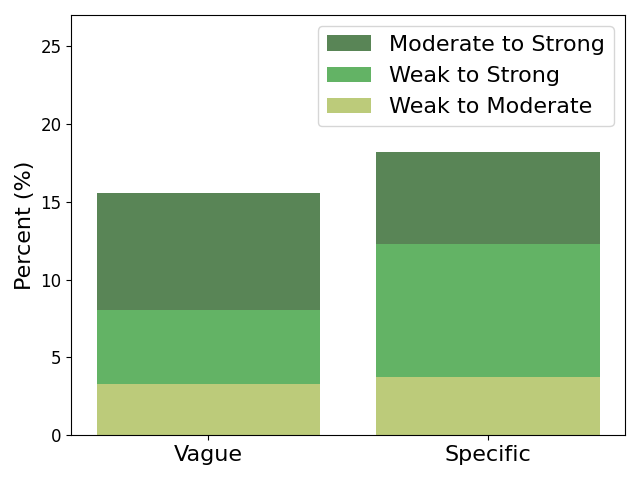}
		\caption{Percentage of users who improved the strength of their password.}
		\label{fig:improvement-specificity}
	\end{center}
\end{figure}

\subsection{Source-dependence Effect}

We found that the specificity of the prompt had no significant effect on password selection.  Of the users who saw a vague prompt, 15.6\% opted to go back and ultimately selected a stronger password, compared to 18.2\% of users who saw the specific prompt. This difference, depicted in Figure~\ref{fig:improvement-specificity}, was not statistically significant ($p=.573$). 

This negative result may be an indication that the source-dependence effect does not apply in the context of password selection decisions. However, it is also possible that the language of our prompts was insufficient to transfer uncertainty-based risk into probability-based risk in a manner that would trigger the source-dependence effect. Finally, it is possible that many of our users simply did not read the prompt, precluding the possibility of observing statistically significant effects due to the source-dependence effect. 

Regardless of the underlying mechanism, these results suggest that utilizing more specific language about the nature of risks due to weak passwords---including notifying users of how long it would take an attacker to crack a password---is not an effective way to nudge users to select stronger passwords.


\begin{figure}[t!]
	\begin{center}
		\includegraphics[width = \columnwidth]{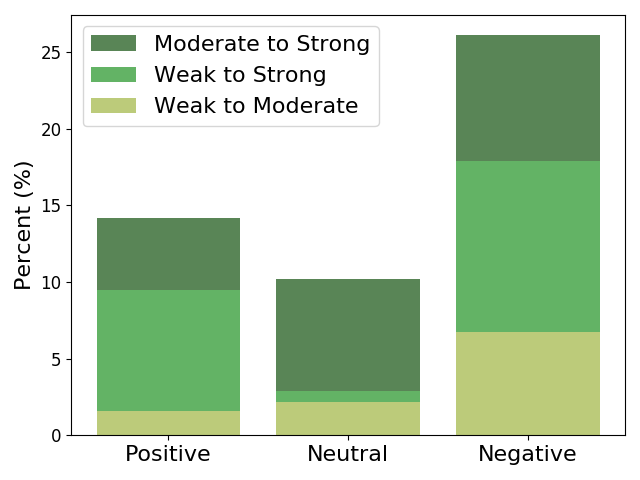}
		\caption{Percentage of users who improved the strength of their password.}
		\label{fig:improvement-framing}
	\end{center}
\end{figure}

\begin{figure}[t!]
	\begin{center}
		\includegraphics[width = \columnwidth]{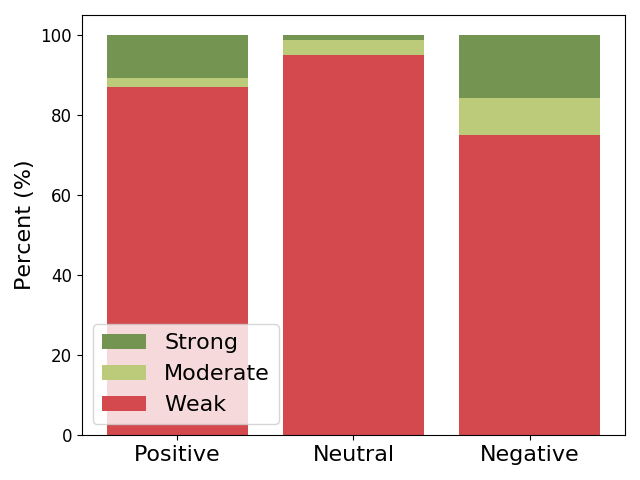}
		\caption{Strength of final password chosen by users who initially chose a weak password.}
		\label{fig:weak_init-framing}
	\end{center}
\end{figure}
	
\subsection{Reference-dependence Effect}

We did find that the framing of the options had a significant effect on password selection decisions.  Approximately 25\% of users improved the strength of their password when presented with the negative framing, a significantly higher rate of improvement than the neutral framing ($p<.001$) or positive framing ($p=.022$). These results are depicted in Figure~\ref{fig:improvement-framing}. 

To confirm that an interaction with negative framing significantly improves password strength, we also tested whether this intervention reduces the number of weak passwords selected by users. Among users assigned to  conditions with negative framing, the number of weak passwords ultimately selected was significantly lower than the number of weak passwords selected initially ($p=.019$). In contrast, the positive and neutral framings showed no significant reduction in the number of weak passwords ultimately selected. 
These results are depicted in Figure~\ref{fig:weak_init-framing}.

These results confirm that the reference-dependence effect occurs in the context of password selection decisions. While further work will be required to validate this result in real-world systems, prior work has found that the results of password studies conducted online generally do extend to real-world systems~\cite{fahl2013ecological}. The insight that reference-dependence effects impact users' password selection decisions therefore provides guidance for how authentication mechanisms designers can prescriptively enhance security: by adding a confirmation page and framing the option to go back and select a strong password as the ``neutral baseline'' (and framing the option to continue with a weak or moderate password as a loss of utility relative to that baseline), we can effectively nudge users to enhance the security of their accounts by selecting significantly stronger passwords.

\subsection{Mental Models}

In our follow-up survey, we asked about users' mental models of hacking in order to explore whether there was a correlation between how users thought about password hacking attacks and how users responded to our interactive prompts.

\begin{figure}[t!]
	\begin{center}
		\includegraphics[width = \columnwidth]{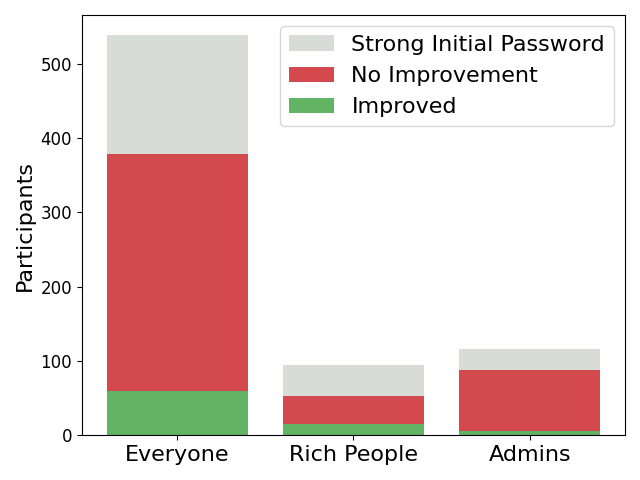}
		\caption{Password selection decisions broken down by perceived target of attacks.}
		\label{fig:targets_corr}
	\end{center}
\end{figure}

\paragraph{Hacking Targets.} We asked users to identify who they thought hackers would target during a password-stealing attack: everyone equally, primarily rich people, or primarily privileged users (e.g., system administrators). We found that 70.7\% of participants believed that hackers target everyone equally, and anyone is equally likely to have their password stolen, 12.3\% of participants believed that hackers primarily targeted rich people, and 15.4\% of participants believed that hackers primarily target privileged accounts (e.g., system administrators). These results are depicted in Figure~\ref{fig:targets_corr}.

A small number of participants opted instead to provide a free-form response. Some of these responses identified the primary targets as ``gullible people'' and ``weak links''. Other responses provided more nuanced variants of the options provided, e.g., ``It depends on the hacker. Botnets attack everyone while social engineering attacks focus on special privleges.'' or identified all of the above as the best description of who is likely to be the target of a password stealing attack. 

Users who believed that hackers primarily target administrators during such attacks were significantly less likely to improve the strength of their password after exposure to the interactive prompt compared to users who believed that everyone is targeted equally ($p=.044$). We believe this difference occurs because users with this mental model are less likely to believe they will be the target of an attack. To our surprise, users who believed that hackers primarily target rich people were actually more likely to improve the strength of their password compared to users who believed that everyone is targeted equally ($p=.040$) and were also significantly more likely to initially select a strong password ($p=.011$). This may be due to the fact that Americans consistently underestimate income inequality~\cite{norton2011building,norton2014not,davidai2018americans} and the income of top earners relative to the median worker~\cite{kiatpongsan2014much}, and thus may consider themselves to be a high-priority target if they hold that mental model.

\begin{figure}[t!]
	\begin{center}
		\includegraphics[width = \columnwidth]{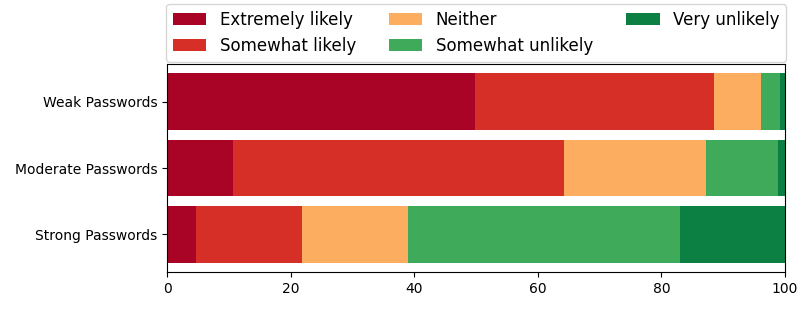}
		\caption{Perceptions of how likely a password guessing attack is to succeed based on the strength of a user's password.}
		\label{fig:risk}
	\end{center}
\end{figure}

\paragraph{Risk Evaluation.} We also asked survey participants to rate how likely a password guessing attack would be to succeed if a user selected a weak password, a moderate password, or a strong password. We found that 88.6\% of participants considered a weak password to be somewhat or very likely to be successfully attacked, compared to 63.9\% of participants for a moderate and 21.8\% of participants for a strong password. These responses, which are depicted in Figure~\ref{fig:risk}, were statistically significantly different between all the different password strengths ($p<.001$). They suggest that most users believe that stronger passwords are in fact less likely to be vulnerable to password guessing attacks. However, there was no significant correlation between whether a user believed stronger passwords had less risk and whether that user improved the strength of their password after seeing the interactive prompt ($p=.873$).

\begin{figure}[t!]
	\begin{center}
		\includegraphics[width = \columnwidth]{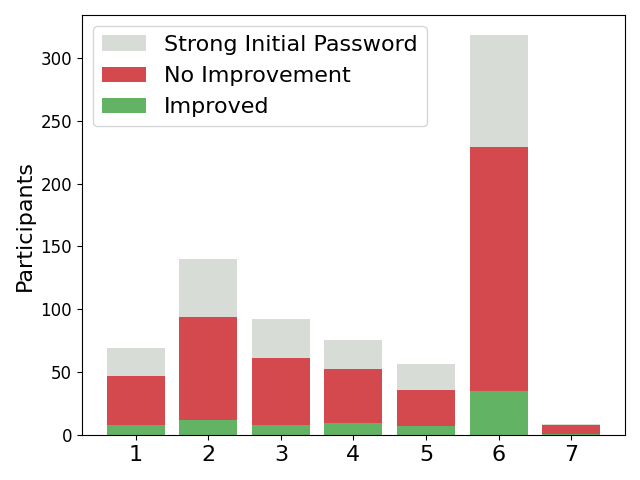}
		\caption{Password selection decisions as a function of number of negative consequences of password compromise.}
		\label{fig:num_dangers_corr}
	\end{center}
\end{figure}

We then asked users to identify what a hacker would be able to do if they successfully learned the user's password. Participants were asked so select all applicable choices from a provided list of six possible consequences; they could also enter a free-form response. We found that 43.0\% of respondents selected all six options from the list of possibilities provided, with nine of those participants also adding a seventh free-form answer. These results, depicted in Figure~\ref{fig:num_dangers_corr}, suggest that many users believe there are lots of potential negative consequences due to (successfully) password hacking attacks. However, there was no significant correlation between the number of consequences a participant selected and how likely they were to improve their password or between the number of consequences they selected and how likely they were to initially select a strong password. 

These negative results suggest that the reference-dependence effect applies uniformly across different risk assessments. They therefore imply that a prescriptive intervention that leverages this effect is likely to enhance security for many different classes of users, regardless of how well-informed they are about the risks associated with compromised passwords.

\section{Discussion and Limitations}

These results suggest that it may be possible to significantly improve the strength of user-selected passwords by leveraging insights from prospect theory---in particular the reference-dependence effect---through a negatively framed interactive prompt after users select an initial password. However, further work will be required to validate these results and to ensure that the potential benefits to security outweigh any potential harms.

\begin{figure}[t!]
	\begin{center}
		\includegraphics[width = \columnwidth]{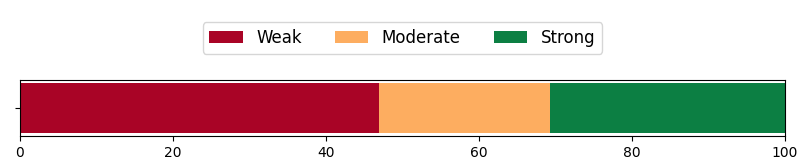}
		\caption{Strength of passwords initially selected by users (prior to seeing the interactive prompt).}
		\label{fig:init_strength}
	\end{center}
\end{figure}

\paragraph{Ecological Validity.} The major limitation of this work arises from the fact that we recruited Amazon Mechanical Turk users to select passwords for an experimental account rather than conducting the study using an authentication system for real accounts. Prior work has found that Mechanical Turk users select slightly weaker passwords in experimental settings compared to users selecting passwords for real accounts. For example, one study found that 44.0\% of users selected guessable passwords for their real account compared to 47.5\% of Mechanical Turk users who were asked to select a password for an experimental study account given identical constraints~\cite{mazurek2013measuring}. In our study, we similarly found that 47.0\% of our users initially selected a weak password (Figure~\ref{fig:init_strength}), so we expect that this number is slightly higher than we would see for real accounts. Nonetheless, prior work has found that results from laboratory and online studies about passwords correspond to patterns in behavior for real accounts~\cite{fahl2013ecological}. While follow-up work will be needed to validate the reference-dependence effect for real accounts and to investigate whether it predicts behavior for different types of accounts (e.g., importand and unimportant accounts), we hypothesize that this effect will extend to  real-world password selection decisions.

\paragraph{Memorability.} The risk of account compromise due to password cracking and other attacks is not the only risk that users consider when selecting a password: users also need to weigh risks associated with other factors such as memorability. Forgetting a password is inconvenient in the best case; in the worst case it can cause users to lose access to accounts. Future work will be required to determine the effect of framing on the memorability of passwords that users select. 

Concerns about memorability might motivate users to employ memory-assistance techniques. This could lead to improved security practices---such as increased adoption of password managers---or to bad security practices---such as writing down passwords and leaving them in accessible locations. Further work will be required to evaluate the impact of framing on these other password-related practices. 

\paragraph{Ethical Considerations.} Leveraging the reference-dependence effect through negative framing of decisions has the potential to enhance security by encouraging users to adopt stronger passwords. However, this effect is an example of nudging~\cite{acquisti2017nudges}. While nudging is often associated with the class of UI design elements known as dark patterns~\cite{conti2010malicious, brignull2019dark, bosch2016tales, gray2018dark}, which nudge users to make decisions that are inimical to their interests, nudging can also be used towards making decisions that the mechanisms designer views as ``better'', a form of nudging sometimes called \emph{soft paternalism}~\cite{nudge,schnellenbach2012nudges,kirchgassner2017soft,fateh2014governing}. Since nudging inherently leverages subconscious patterns in human behavior, care and consideration will be required to ensure that any presecriptive application of nudging and prospect theory effects with real-world impact---including leveraging the referenced-dependence effect to improve password selection---is handled ethically and responsibly.

\section{Related Work}


\paragraph{Improving Password Selections.} Given the prevalence of password-based authentication, a large body of work has been dedicated to the problem of improving users' password selection decisions. 

Early work on estimating password strength generally focused on entropy-based metrics~\cite{komanduri2011passwords}. However, entropy has since been criticized as been a poor measure of password guessability~\cite{weir2010testing,kelley2012guess,wheeler2016zxcvbn}. More recent efforts use dictionaries of words and leaked passwords and/or variants of words in those dictionaries (e.g., L33t-style substitutions) to define classes of weak or prohibited passwords~\cite{wheeler2016zxcvbn,golla2018accuracy,mazurek2013measuring}. 

Studies have found that users exhibit misconceptions about password strength~\cite{ur2015added}, which has resulting in increasing adoption of password meters across the most popular websites~\cite{egelman2013does}. In general, having a password meter improves password strength, especially for accounts that users consider important~\cite{egelman2013does}. However, some websites continue to use metrics that rely on entropy-based metrics and are thus ineffective and enforcing strong password selections~\cite{weir2010testing}; one study found that most password meters on websites are ineffective~\cite{carnavalet2015large}. Careful calibration is also required to ensure that usability considerations don't undermine the benefits of a password meter; meters that are too strict can annoy users, while meters that are too lenient fail to exhibit the benefits of stronger password selections~\cite{ur2012does}.

\paragraph{Applications of Prospect Theory to Security.} Despite the success of prospect theory in economics, there has been limited work applying prospect theory to security decisions, and only in limited domains. Verendel~\cite{verendel2008prospect} developed a prospect theory model for decisions about buying versus skipping security protections (e.g., anti-virus software), although that work did not include any experimental validation. Schroeder~\cite{schroeder2005using} conducted a lab-based survey of IT officers in the U.S. military and found that prospect theory predicted hypothetical decisions about investment in information security.  Sawicka and Gonzalez~\cite{sawicka2003choice} explored the extent to which prospect theory can explain behavioral dynamics in IT-based work environments; they found the model matched choices observed in a short experimental run, but that it was not likely to account accurately for behavior over longer time periods. Sanjab et al.~\cite{sanjab2017prospect} explored how the decision weight function and value function impact principals' decisions in adversarial games in the context of attacks on Unmanned Aerial Vehicles (UAVs); they found that these subjective functions led to the adoption of riskier strategies which cause delays in delivery. Most recently, Qu et al.~\cite{qu2019towards} investigated the reference-dependence effect and the pseudocertainty effect in the context of two-factor authentication; they found that both effects explained whether or not users choose to enable two-factor authentication for a game in a laboratory setting. However, other security decisions---notably including password selection---have not been previously studied.

\paragraph{Applications of Prospect Theory to Privacy.} In 2007, Acquisti et al. posited that several prospect theory effects---notably ambiguity aversion---might significantly impact privacy decision making~\cite{acquisti2007can}. Follow-up work found that people were more willing to sell personal information than to buy back previously-disclosed information~\cite{grossklags200725,acquisti2013privacy}, and that the framing of notice affected whether or not users disclosed personal information in a survey~\cite{adjerid2013sleights}. Chloe et al.~\cite{choe2013nudging} also found that visual signals of an app's trustworthiness were affected by framing, with positively framed signals proving more effective at influencing user opinions about the trustworthiness of an app. More recent work has looked at developing and validating a theory for how context and personality affect decisions about disclosing personal information~\cite{bansal2016context} and at the mechanism-design problem of how to calibrate noise in privacy-preserving mechanisms~\cite{liao2017optimal,liao2019prospect}. 

\section{Conclusion}

In this work, we explore the extent to which two effects identified in the prospect theory literature---the source-dependence effect and the reference-dependence effect---apply to users' password selection decisions, and we evaluate the feasibility of leveraging these effects to enhance security. We conduct a user study with 762 users to explore the impact of these two effects. Although source-dependence has no significant effect on password selection,  we find that the reference-dependence effect does. By employing a negative framing in a follow-up interactive prompt, we can nudge 25\% of users to increase the strength of their selected password. This effect appears to be consistent across different mental models of password risks. 

Further work will be required to validate the reference-dependence effect for real-world accounts, and careful consideration will be required to ensure that framing and any other nudging effects are employed ethically. Nonetheless, these results suggest a path forward that will enhance system security by nudging users to adopt significantly stronger passwords. 

\newpage
\bibliographystyle{plain}
\bibliography{main}

\begin{thebibliography}{10}

\bibitem{acquisti2017nudges}
Alessandro Acquisti, Idris Adjerid, Rebecca Balebako, Laura Brandimarte,
  Lorrie~Faith Cranor, Saranga Komanduri, Pedro~Giovanni Leon, Norman Sadeh,
  Florian Schaub, Manya Sleeper, et~al.
\newblock Nudges for privacy and security: Understanding and assisting users’
  choices online.
\newblock {\em ACM Computing Surveys (CSUR)}, 50(3):1--41, 2017.

\bibitem{acquisti2007can}
Alessandro Acquisti, Stefanos Gritzalis, Costos Lambrinoudakis, and Sabrina
  di~Vimercati.
\newblock {\em What can behavioral economics teach us about privacy?}
\newblock Auerbach Publications, 2007.

\bibitem{acquisti2013privacy}
Alessandro Acquisti, Leslie~K John, and George Loewenstein.
\newblock What is privacy worth?
\newblock {\em The Journal of Legal Studies}, 42(2):249--274, 2013.

\bibitem{adjerid2013sleights}
Idris Adjerid, Alessandro Acquisti, Laura Brandimarte, and George Loewenstein.
\newblock Sleights of privacy: Framing, disclosures, and the limits of
  transparency.
\newblock In {\em Proceedings of the ninth symposium on usable privacy and
  security}, pages 1--11, 2013.

\bibitem{bansal2016context}
Gaurav Bansal, Fatemeh~Mariam Zahedi, and David Gefen.
\newblock Do context and personality matter? trust and privacy concerns in
  disclosing private information online.
\newblock {\em Information \& Management}, 53(1):1--21, 2016.

\bibitem{barberis2008stocks}
Nicholas Barberis and Ming Huang.
\newblock Stocks as lotteries: The implications of probability weighting for
  security prices.
\newblock {\em American Economic Review}, 98(5):2066--2100, 2008.

\bibitem{barseghyan2013nature}
Levon Barseghyan, Francesca Molinari, Ted O'Donoghue, and Joshua~C Teitelbaum.
\newblock The nature of risk preferences: Evidence from insurance choices.
\newblock {\em American Economic Review}, 103(6):2499--2529, 2013.

\bibitem{bosch2016tales}
Christoph B{\"o}sch, Benjamin Erb, Frank Kargl, Henning Kopp, and Stefan
  Pfattheicher.
\newblock Tales from the dark side: Privacy dark strategies and privacy dark
  patterns.
\newblock {\em Proc. Priv. Enhancing Technol.}, 2016(4):237--254, 2016.

\bibitem{brignull2019dark}
Harry Brignull and Alexander Darlo.
\newblock Dark patterns.
\newblock {\em Dark Patterns}, 2019.

\bibitem{camerer1997labor}
Colin Camerer, Linda Babcock, George Loewenstein, and Richard Thaler.
\newblock Labor supply of {N}ew {Y}ork {C}ity cabdrivers: One day at a time.
\newblock {\em The Quarterly Journal of Economics}, 112(2):407--441, 1997.

\bibitem{carnavalet2015large}
Xavier De Carn{\'e}~De Carnavalet and Mohammad Mannan.
\newblock A large-scale evaluation of high-impact password strength meters.
\newblock {\em ACM Transactions on Information and System Security (TISSEC)},
  18(1):1--32, 2015.

\bibitem{choe2013nudging}
Eun~Kyoung Choe, Jaeyeon Jung, Bongshin Lee, and Kristie Fisher.
\newblock Nudging people away from privacy-invasive mobile apps through visual
  framing.
\newblock In {\em IFIP Conference on Human-Computer Interaction}, pages 74--91.
  Springer, 2013.

\bibitem{conti2010malicious}
Gregory Conti and Edward Sobiesk.
\newblock Malicious interface design: exploiting the user.
\newblock In {\em Proceedings of the 19th international conference on World
  wide web}, pages 271--280, 2010.

\bibitem{crawford2011new}
Vincent~P Crawford and Juanjuan Meng.
\newblock New york city cab drivers' labor supply revisited:
  Reference-dependent preferences with rational-expectations targets for hours
  and income.
\newblock {\em American Economic Review}, 101(5):1912--32, 2011.

\bibitem{davidai2018americans}
Shai Davidai.
\newblock Why do americans believe in economic mobility? economic inequality,
  external attributions of wealth and poverty, and the belief in economic
  mobility.
\newblock {\em Journal of Experimental Social Psychology}, 79:138--148, 2018.

\bibitem{dimmock2010loss}
Stephen~G Dimmock and Roy Kouwenberg.
\newblock Loss-aversion and household portfolio choice.
\newblock {\em Journal of Empirical Finance}, 17(3):441--459, 2010.

\bibitem{egelman2013does}
Serge Egelman, Andreas Sotirakopoulos, Ildar Muslukhov, Konstantin Beznosov,
  and Cormac Herley.
\newblock Does my password go up to eleven? {T}he impact of password meters on
  password selection.
\newblock In {\em Proceedings of the SIGCHI Conference on Human Factors in
  Computing Systems}, CHI '13, pages 2379--2388, New York, NY, USA, 2013.
  Association for Computing Machinery.

\bibitem{fahl2013ecological}
Sascha Fahl, Marian Harbach, Yasemin Acar, and Matthew Smith.
\newblock On the ecological validity of a password study.
\newblock In {\em Proceedings of the Ninth Symposium on Usable Privacy and
  Security}, pages 1--13, 2013.

\bibitem{fateh2014governing}
Bijan Fateh-Moghadam and Thomas Gutmann.
\newblock Governing [through] autonomy. the moral and legal limits of “soft
  paternalism”.
\newblock {\em Ethical Theory and Moral Practice}, 17(3):383--397, 2014.

\bibitem{friedman1948utility}
Milton Friedman and Leonard~J Savage.
\newblock The utility analysis of choices involving risk.
\newblock {\em Journal of political Economy}, 56(4):279--304, 1948.

\bibitem{fryer2012enhancing}
Roland~G Fryer~Jr, Steven~D Levitt, John List, and Sally Sadoff.
\newblock Enhancing the efficacy of teacher incentives through loss aversion: A
  field experiment.
\newblock Technical report, National Bureau of Economic Research, 2012.

\bibitem{golla2018accuracy}
Maximilian Golla and Markus D{\"u}rmuth.
\newblock On the accuracy of password strength meters.
\newblock In {\em Proceedings of the 2018 ACM SIGSAC Conference on Computer and
  Communications Security}, pages 1567--1582, 2018.

\bibitem{gray2018dark}
Colin~M Gray, Yubo Kou, Bryan Battles, Joseph Hoggatt, and Austin~L Toombs.
\newblock The dark (patterns) side of {UX} design.
\newblock In {\em Proceedings of the 2018 CHI Conference on Human Factors in
  Computing Systems}, pages 1--14, 2018.

\bibitem{grossklags200725}
Jens Grossklags and Alessandro Acquisti.
\newblock When 25 cents is too much: An experiment on willingness-to-sell and
  willingness-to-protect personal information.
\newblock In {\em WEIS}, 2007.

\bibitem{heidhues2014regular}
Paul Heidhues and Botond K{\H{o}}szegi.
\newblock Regular prices and sales.
\newblock {\em Theoretical Economics}, 9(1):217--251, 2014.

\bibitem{hossain2012behavioralist}
Tanjim Hossain and John~A List.
\newblock The behavioralist visits the factory: Increasing productivity using
  simple framing manipulations.
\newblock {\em Management Science}, 58(12):2151--2167, 2012.

\bibitem{hu2007behavioral}
Wei-Yin Hu and Jason~S Scott.
\newblock Behavioral obstacles in the annuity market.
\newblock {\em Financial Analysts Journal}, 63(6):71--82, 2007.

\bibitem{kahneman1979prospect}
Daniel Kahneman and Amos Tversky.
\newblock Prospect theory: An analysis of decision under risk.
\newblock {\em Econometrica}, 47(2):263--292, 1979.

\bibitem{kelley2012guess}
Patrick~Gage Kelley, Saranga Komanduri, Michelle~L Mazurek, Richard Shay,
  Timothy Vidas, Lujo Bauer, Nicolas Christin, Lorrie~Faith Cranor, and Julio
  Lopez.
\newblock Guess again (and again and again): Measuring password strength by
  simulating password-cracking algorithms.
\newblock In {\em 2012 IEEE symposium on security and privacy}, pages 523--537.
  IEEE, 2012.

\bibitem{kiatpongsan2014much}
Sorapop Kiatpongsan and Michael~I Norton.
\newblock How much (more) should ceos make? a universal desire for more equal
  pay.
\newblock {\em Perspectives on Psychological Science}, 9(6):587--593, 2014.

\bibitem{kirchgassner2017soft}
Gebhard Kirchg{\"a}ssner.
\newblock Soft paternalism, merit goods, and normative individualism.
\newblock {\em European Journal of Law and Economics}, 43(1):125--152, 2017.

\bibitem{komanduri2011passwords}
Saranga Komanduri, Richard Shay, Patrick~Gage Kelley, Michelle~L Mazurek, Lujo
  Bauer, Nicolas Christin, Lorrie~Faith Cranor, and Serge Egelman.
\newblock Of passwords and people: {M}easuring the effect of
  password-composition policies.
\newblock In {\em Proceedings of the sigchi conference on human factors in
  computing systems}, pages 2595--2604, 2011.

\bibitem{kHoszegi2007reference}
Botond K{\H{o}}szegi and Matthew Rabin.
\newblock Reference-dependent risk attitudes.
\newblock {\em American Economic Review}, 97(4):1047--1073, 2007.

\bibitem{kHoszegi2009reference}
Botond K{\H{o}}szegi and Matthew Rabin.
\newblock Reference-dependent consumption plans.
\newblock {\em American Economic Review}, 99(3):909--36, 2009.

\bibitem{levitt2016behavioralist}
Steven~D Levitt, John~A List, Susanne Neckermann, and Sally Sadoff.
\newblock The behavioralist goes to school: Leveraging behavioral economics to
  improve educational performance.
\newblock {\em American Economic Journal: Economic Policy}, 8(4):183--219,
  2016.

\bibitem{liao2017optimal}
Guocheng Liao, Xu~Chen, and Jianwei Huang.
\newblock Optimal privacy-preserving data collection: A prospect theory
  perspective.
\newblock In {\em GLOBECOM 2017-2017 IEEE Global Communications Conference},
  pages 1--6. IEEE, 2017.

\bibitem{liao2019prospect}
Guocheng Liao, Xu~Chen, and Jianwei Huang.
\newblock Prospect theoretic analysis of privacy-preserving mechanism.
\newblock {\em IEEE/ACM Transactions on Networking}, 28(1):71--83, 2019.

\bibitem{mazurek2013measuring}
Michelle~L Mazurek, Saranga Komanduri, Timothy Vidas, Lujo Bauer, Nicolas
  Christin, Lorrie~Faith Cranor, Patrick~Gage Kelley, Richard Shay, and Blase
  Ur.
\newblock Measuring password guessability for an entire university.
\newblock In {\em Proceedings of the 2013 ACM SIGSAC conference on Computer \&
  communications security}, pages 173--186, 2013.

\bibitem{meng2018can}
Juanjuan Meng and Xi~Weng.
\newblock Can prospect theory explain the disposition effect? {A} new
  perspective on reference points.
\newblock {\em Management Science}, 64(7):3331--3351, 2018.

\bibitem{passwords2020}
NordPass.
\newblock Top 200 most common passwords of the year 2020. (access october 11,
  2021).
\newblock https://nordpass.com/most-common-passwords-list/.

\bibitem{norton2011building}
Michael~I Norton and Dan Ariely.
\newblock Building a better america—one wealth quintile at a time.
\newblock {\em Perspectives on psychological science}, 6(1):9--12, 2011.

\bibitem{norton2014not}
Michael~Irwin Norton, David~T Neal, Cassandra~L Govan, Dan Ariely, and Elise
  Holland.
\newblock The not-so-common-wealth of australia: Evidence for a cross-cultural
  desire for a more equal distribution of wealth.
\newblock {\em Analyses of Social Issues and Public Policy}, 2014.

\bibitem{qu2019towards}
Leilei Qu, Cheng Wang, Ruojin Xiao, Jianwei Hou, Wenchang Shi, and Bin Liang.
\newblock Towards better security decisions: {A}pplying prospect theory to
  cybersecurity.
\newblock In {\em Extended Abstracts of the 2019 CHI Conference on Human
  Factors in Computing Systems}, pages 1--6, 2019.

\bibitem{redmiles2019well}
Elissa~M Redmiles, Sean Kross, and Michelle~L Mazurek.
\newblock How well do my results generalize? comparing security and privacy
  survey results from mturk, web, and telephone samples.
\newblock In {\em 2019 IEEE Symposium on Security and Privacy (SP)}, pages
  1326--1343. IEEE, 2019.

\bibitem{sanjab2017prospect}
Anibal Sanjab, Walid Saad, and Tamer Ba{\c{s}}ar.
\newblock Prospect theory for enhanced cyber-physical security of drone
  delivery systems: A network interdiction game.
\newblock In {\em 2017 IEEE international conference on communications (ICC)},
  pages 1--6. IEEE, 2017.

\bibitem{sawicka2003choice}
Agata Sawicka and Jose~J Gonzalez.
\newblock Choice under risk in {IT}-environments according to cumulative
  prospect theory.
\newblock In {\em 21st International Conference of the System Dynamics Society,
  New York}, 2003.

\bibitem{schnellenbach2012nudges}
Jan Schnellenbach.
\newblock Nudges and norms: On the political economy of soft paternalism.
\newblock {\em European Journal of Political Economy}, 28(2):266--277, 2012.

\bibitem{schroeder2005using}
Neil~J Schroeder.
\newblock Using prospect theory to investigate decision-making bias within an
  information security context.
\newblock Technical report, Air Force Institution of Technology
  Wright-Patterson School of Engineering and Management, 2005.

\bibitem{shefrin1985disposition}
Hersh Shefrin and Meir Statman.
\newblock The disposition to sell winners too early and ride losers too long:
  Theory and evidence.
\newblock {\em The Journal of finance}, 40(3):777--790, 1985.

\bibitem{snowberg2010explaining}
Erik Snowberg and Justin Wolfers.
\newblock Explaining the favorite--long shot bias: Is it risk-love or
  misperceptions?
\newblock {\em Journal of Political Economy}, 118(4):723--746, 2010.

\bibitem{sydnor2010over}
Justin Sydnor.
\newblock ({O}ver) insuring modest risks.
\newblock {\em American Economic Journal: Applied Economics}, 2(4):177--99,
  2010.

\bibitem{nudge}
Richard Thaler and Cass Sunstein.
\newblock {\em Nudge: Improving Decisions About Health, Wealth, and Happiness}.
\newblock Yale University Press, 2008.

\bibitem{thaler2004save}
Richard~H Thaler and Shlomo Benartzi.
\newblock Save more tomorrow: Using behavioral economics to increase employee
  saving.
\newblock {\em Journal of political Economy}, 112(S1):S164--S187, 2004.

\bibitem{tversky1981framing}
Amos Tversky and Daniel Kahneman.
\newblock The framing of decisions and the psychology of choice.
\newblock {\em science}, 211(4481):453--458, 1981.

\bibitem{tversky1986framing}
Amos Tversky and Daniel Kahneman.
\newblock The framing of decisions and the evaluation of prospects.
\newblock In {\em Studies in Logic and the Foundations of Mathematics}, volume
  114, pages 503--520. Elsevier, 1986.

\bibitem{tversky1991loss}
Amos Tversky and Daniel Kahneman.
\newblock Loss aversion in riskless choice: A reference-dependent model.
\newblock {\em The quarterly journal of economics}, 106(4):1039--1061, 1991.

\bibitem{tversky1992advances}
Amos Tversky and Daniel Kahneman.
\newblock Advances in prospect theory: Cumulative representation of
  uncertainty.
\newblock {\em Journal of Risk and uncertainty}, 5(4):297--323, 1992.

\bibitem{ur2012does}
Blase Ur, Patrick~Gage Kelley, Saranga Komanduri, Joel Lee, Michael Maass,
  Michelle~L Mazurek, Timothy Passaro, Richard Shay, Timothy Vidas, Lujo Bauer,
  et~al.
\newblock How does your password measure up? {T}he effect of strength meters on
  password creation.
\newblock In {\em 21st {USENIX} Security Symposium}, pages 65--80, 2012.

\bibitem{ur2015added}
Blase Ur, Fumiko Noma, Jonathan Bees, Sean~M Segreti, Richard Shay, Lujo Bauer,
  Nicolas Christin, and Lorrie~Faith Cranor.
\newblock " i added'!'at the end to make it secure": Observing password
  creation in the lab.
\newblock In {\em Eleventh Symposium On Usable Privacy and Security
  ($\{$SOUPS$\}$ 2015)}, pages 123--140, 2015.

\bibitem{verendel2008prospect}
Vilhelm Verendel.
\newblock {\em A prospect theory approach to security}.
\newblock Citeseer, 2008.

\bibitem{vonNeumann1944theory}
John von Neumann and Oskar Morgenstern.
\newblock {\em Theory of Games and Economic Behavior}.
\newblock Princeton University Press, 1944.

\bibitem{wash2010folk}
Rick Wash.
\newblock Folk models of home computer security.
\newblock In {\em Proceedings of the Sixth Symposium on Usable Privacy and
  Security}, pages 1--16, 2010.

\bibitem{weir2010testing}
Matt Weir, Sudhir Aggarwal, Michael Collins, and Henry Stern.
\newblock Testing metrics for password creation policies by attacking large
  sets of revealed passwords.
\newblock In {\em Proceedings of the 17th ACM conference on Computer and
  communications security}, pages 162--175, 2010.

\bibitem{wheeler2016zxcvbn}
Daniel~Lowe Wheeler.
\newblock zxcvbn: Low-budget password strength estimation.
\newblock In {\em 25th {USENIX} Security Symposium}, pages 157--173, 2016.

\end{thebibliography}

\newpage
\appendix
\section{Survey Questions}\label{appendix:survey}

\begin{enumerate}
	\item How strong was the password you chose when you created your account on the site?
	\begin{itemize}
		\item Strong
		\item Moderate
		\item Weak
	\end{itemize}
	\item How much do you agree with the statement: A hacker would be likely to try to hack this site. 
	\begin{itemize}
		\item Completely agree
		\item Somewhat agree
		\item Neither agree nor disagree
		\item Somewhat disagree
		\item Completely disagree
	\end{itemize}
	\item How much do you agree with the statement: A hacker would be likely to successfully guess the password I used on this site.
	\begin{itemize}
		\item Completely agree
		\item Somewhat agree
		\item Neither agree nor disagree
		\item Somewhat disagree
		\item Completely disagree
	\end{itemize}
	\item Is the password you used on this site a password that you also use on other sites?
	\begin{itemize}
		\item Yes
		\item No
	\end{itemize}
	\item How common are password stealing attacks?
		\begin{itemize}
			\item Extremely common
			\item Somewhat common
			\item Neither common nor uncommon
			\item Somewhat uncommon
			\item Extremely uncommon
		\end{itemize}
	\item How could hackers potentially learn your password? Choose all that apply.
	\begin{itemize}
		\item It is impossible for a hacker to learn my password.
		\item If I accidentally download a virus, a malicious app, or a malicious attachment.
		\item If I visit a sketchy or malicious website.
		\item If I accidentally click on a phishing link and enter my credentials on a fake website.
		\item If a hacker (or a program run by a hacker) guesses my password on the website.
		\item If a hacker steals the files storing all passwords for the website.
		\item Other: \underline{\hspace{120pt}}
	\end{itemize}
	\item How likely would it be for a password stealing attack to succeed if you use a weak password?
			\begin{itemize}
				\item Extremely likely
				\item Somewhat likely
				\item Neither likely nor unlikely
				\item Somewhat unlikely
				\item Extremely unlikely
			\end{itemize}
	\item How likely would it be for a password stealing attack to succeed if you use a moderate password?
			\begin{itemize}
				\item Extremely likely
				\item Somewhat likely
				\item Neither likely nor unlikely
				\item Somewhat unlikely
				\item Extremely unlikely
			\end{itemize}
	\item How likely would it be for a password stealing attack to succeed if you use a strong password?
			\begin{itemize}
				\item Extremely likely
				\item Somewhat likely
				\item Neither likely nor unlikely
				\item Somewhat unlikely
				\item Extremely unlikely
			\end{itemize}
	\item Do you think upgrading your passwords can prevent password guessing?  
	\begin{itemize}
		\item Yes
		\item Maybe
		\item No
	\end{itemize}
	\item What could a hacker do if they successfully learn your password? Choose all that apply.
	\begin{itemize}
		\item They could cause bugs (viruses can cause computers to crash, quit applications, erase important system files).
		\item They could steal personal and financial information from individual computers, and send the information to criminal.
		\item They could resell personal information.
		\item They could display annoying visual images on computers (a skull, advertising popups, or pornography).
		\item They could control the computer and use the computer to send information to others.
		\item They could use the computers to cause problems for third parties.
		\item Other: \underline{\hspace{120pt}}
	\end{itemize}
	\item Which of the following are likely to try to steal passwords? Choose all that apply.
	\begin{itemize}
		\item A young computer geek who wants to show off or explore the internet
		\item Criminals
		\item Organizations and institutions
		\item Other: \underline{\hspace{120pt}}
	\end{itemize}
	\item Which of the following best describes who is likely to be the target of a password stealing attack?
	\begin{itemize}
		\item Hackers target everyone equally, and anyone is equally likely to have their password stolen
		\item Hackers primarily target rich people
		\item Hackers primarily target people with special privileges (e.g, system administrators)
		\item Other: \underline{\hspace{120pt}}
	\end{itemize}
	\item What is your current age?
	\begin{itemize}
		\item 18-24
		\item 25-34
		\item 35-44
		\item 45-59
		\item 60-74
		\item 75+
	\end{itemize}
	\item What is your gender?
	\begin{itemize}
		\item Man
		\item Woman
		\item Non-binary person
		\item Other: \underline{\hspace{120pt}}
	\end{itemize}
	\item Choose one or more races that you consider yourself to be:
	\begin{itemize}
		\item White
		\item Black or African American
		\item American Indian or Alaska Native
		\item Asian
		\item Pacific Islander or Native Hawaiian
		\item Other: \underline{\hspace{120pt}}
	\end{itemize}
	\item Do you consider yourself to be Hispanic?
	\begin{itemize}
		\item Yes
		\item No
	\end{itemize}
\end{enumerate}

\end{document}